\documentclass[fleqn,usenatbib]{mnras}

\usepackage{amssymb,amsmath,verbatim,mathtools,needspace,enumitem,etoolbox,graphicx,physics,microtype,afterpage,xspace,tabularx,lmodern,multirow, xcolor}
\usepackage{gensymb, bm}
\usepackage{caption}
\usepackage{subcaption}
\usepackage{titlesec}

\titlespacing\section{0pt}{16pt plus 4pt minus 2pt}{4pt plus 2pt minus 2pt}
\titlespacing\subsection{0pt}{12pt plus 4pt minus 2pt}{2pt plus 2pt minus 2pt}
\titlespacing\subsubsection{0pt}{12pt plus 4pt minus 2pt}{0pt plus 2pt minus 2pt}

\usepackage[T1]{fontenc}

\DeclareRobustCommand{\VAN}[3]{#2}
\let\VANthebibliography\thebibliography
\def\thebibliography{\DeclareRobustCommand{\VAN}[3]{##3}\VANthebibliography}

\usepackage{graphicx}	%
\usepackage{amsmath}	%
\usepackage{amssymb}	%
\usepackage{newtxtext,newtxmath}

\newcommand{\revision}[1]{\textcolor{black}{#1}}

\title{Calibration of neutron star natal kick velocities to isolated pulsar observations}

\author[Kapil et al.]{
Veome Kapil,$^{1}$ \thanks{vkapil1@jhu.edu}
Ilya Mandel,$^{2,3}$ \thanks{ilya.mandel@monash.edu}
Emanuele Berti,$^{1}$\thanks{berti@jhu.edu}
Bernhard M\"{u}ller$^2$
\thanks{bernhard.mueller@monash.edu}
\\
$^{1}$Department of Physics and Astronomy, Johns Hopkins University, 3400 N. Charles Street, Baltimore, Maryland, 21218, USA\\
$^{2}$Monash Centre for Astrophysics, School of Physics and Astronomy, Monash University, Clayton, Victoria 3800, Australia\\
$^{3}$The ARC Center of Excellence for Gravitational Wave Discovery -- OzGrav, Australia
}

\pubyear{2022}

\begin{document}
\label{firstpage}
\pagerange{\pageref{firstpage}--\pageref{lastpage}}
\maketitle

\begin{abstract}
Current prescriptions for supernova natal kicks in rapid binary population synthesis simulations are based on fits of simple functions to single pulsar velocity data. We explore a new parameterization of natal kicks received by neutron stars in isolated and binary systems developed by Mandel \& M\"uller, which is based on 1D models and 3D supernova simulations, and accounts for the physical correlations between progenitor properties, remnant mass, and the kick velocity. We constrain two free parameters in this model using very long baseline interferometry velocity measurements of Galactic single pulsars.  We find that the inferred values of natal kick parameters do not differ significantly between single and binary evolution scenarios. The best-fit values of these parameters are $v_{\rm ns} = 520$~km\,s$^{-1}$ for the scaling pre-factor for neutron star kicks, and $\sigma_{\rm ns}=0.3$ for the fractional stochastic scatter in the kick velocities.  %
\end{abstract}

\begin{keywords}
stars: neutron -- supernovae: general -- stars: evolution -- (transients:) neutron star mergers
\end{keywords}

\section{Introduction}\label{sec:intro}

Neutron stars (NSs) form in the core collapse of stars with initial masses between approximately 8 and 20 $M_\odot$ \citep{Woosley:2002zz}.  The associated supernova explosions eject matter at velocities of $\sim 10,000$ km s$^{-1}$ with a significant degree of asymmetry.  Conservation of momentum implies that NSs are born with significant ``natal'' kicks, which are indeed observed among Galactic radio pulsars \citep[e.g.][]{Arzoumanian:2002, Hobbs200510.1111/j.1365-2966.2005.09087.x, FGKaspi:2006, Verbunt:2017zqi, Igoshev:2020}.  These kicks can eject NSs from host environments such as globular clusters, or disrupt stellar binaries if the NS progenitor is a member of a binary.  Consequently, the natal kick prescription used in models such as rapid binary population synthesis codes can significantly affect predictions for the outcomes of such simulations, including for merging double compact objects (DCOs) that may be observable as gravitational-wave sources \citep[e.g.,][]{Broekgaarden:2022}.

There are several challenges associated with the detailed modelling of core-collapse supernovae (CCSNe; \citealt{Janka:2012wk, Muller:2020ard}). In most binary evolution codes, therefore, supernova remnant masses and kicks are typically based on simplified analytical \revision{recipes} \citep{Hurley:2000pk, Fryer:2011cx}, or sampled randomly from fits to the observed velocities of single pulsars \citep{Hobbs200510.1111/j.1365-2966.2005.09087.x, Verbunt:2017zqi}.  One notable exception is the model of \citet{BrayEldridge:2016,BrayEldridge:2018}, which is a phenomenological \revision{model where each natal kick is determined} on the basis of momentum conservation \revision{of the supernova remnant}; this fit was recently calibrated to pulsar observations \revision{and other observed NS populations} by \citet{Richards:2022}.

Recently, \citet{Mandel:2020qwb} (hereafter, MM20) used findings from 3D supernova simulations and 1D models to propose a new parameterized model for computing remnant masses and velocities from the pre-supernova Carbon-Oxygen (CO) core mass. Their supernova kick prescription has a few distinct advantages over the commonly used kick distributions. 
The MM20 prescription is based on physical models and connects the mass of the material ejected from the CO core, which is expected to be tightly coupled to the explosion mechanism, to the amount of  ejected asymmetric linear momentum prior to applying momentum conservation, as in \citet{BrayEldridge:2016}.  This allows for the resulting kick prescription to capture more information about NS-specific properties than population-level fits where all NS kicks are assumed to be drawn from a single distribution. At the same time, the MM20 model accounts for a degree of intrinsic stochasticity in explodability and the explosion mechanism by providing a probabilistic rather than deterministic prescription.  

The MM20 model for NS kicks can be parameterized by just two parameters: $v_{\rm ns}$, a scaling pre-factor for NS kicks, and $\sigma_{\rm kick}$ (hereafter, $\sigma_{\rm ns}$ because we limit our discussion to NS kicks), a measure of the scatter in the kick distribution. The kick received by a NS of mass $M_{\rm ns}$ formed from a progenitor with CO core mass $M_{\rm CO}$ is sampled from a Gaussian distribution with mean 
\begin{equation}
\mu_{\rm kick} = v_{\rm ns} \frac{M_{\rm CO} - M_{\rm ns}}{M_{\rm ns}},
\end{equation}
and standard deviation $\sigma_{\rm ns} \mu_{\rm kick}$. With only two parameters to tune, the model mitigates the risk of over-fitting to observational data; in fact, the reproducibility of observations can be viewed as a test of the model.

In this paper, we use observational data of single pulsar velocities to constrain these two parameters for NS natal kicks and study some of the properties of the resulting kick distributions. The rest of this paper is organized as follows. In Section~\ref{sec:observations}, we briefly introduce the pulsar velocity data which will be used to constrain the MM20 model parameters ($v_{\rm ns}$ and $\sigma_{\rm ns}$). In Section~\ref{sec:sse}, we apply these constraints to the scenario where all observed pulsars originate from single stellar evolution. In Section~\ref{sec:bse}, we explore the more realistic scenario where all the pulsars originate in binary systems. In Section~\ref{sec:discussion}, we discuss the impacts of the MM20 natal kick model on local detection rates for binary NSs (BNSs), as well as NS retention rates in globular clusters.

\section{Pulsar Velocity Observations}\label{sec:observations}
\begin{figure}
\includegraphics[clip,width=1.\linewidth]{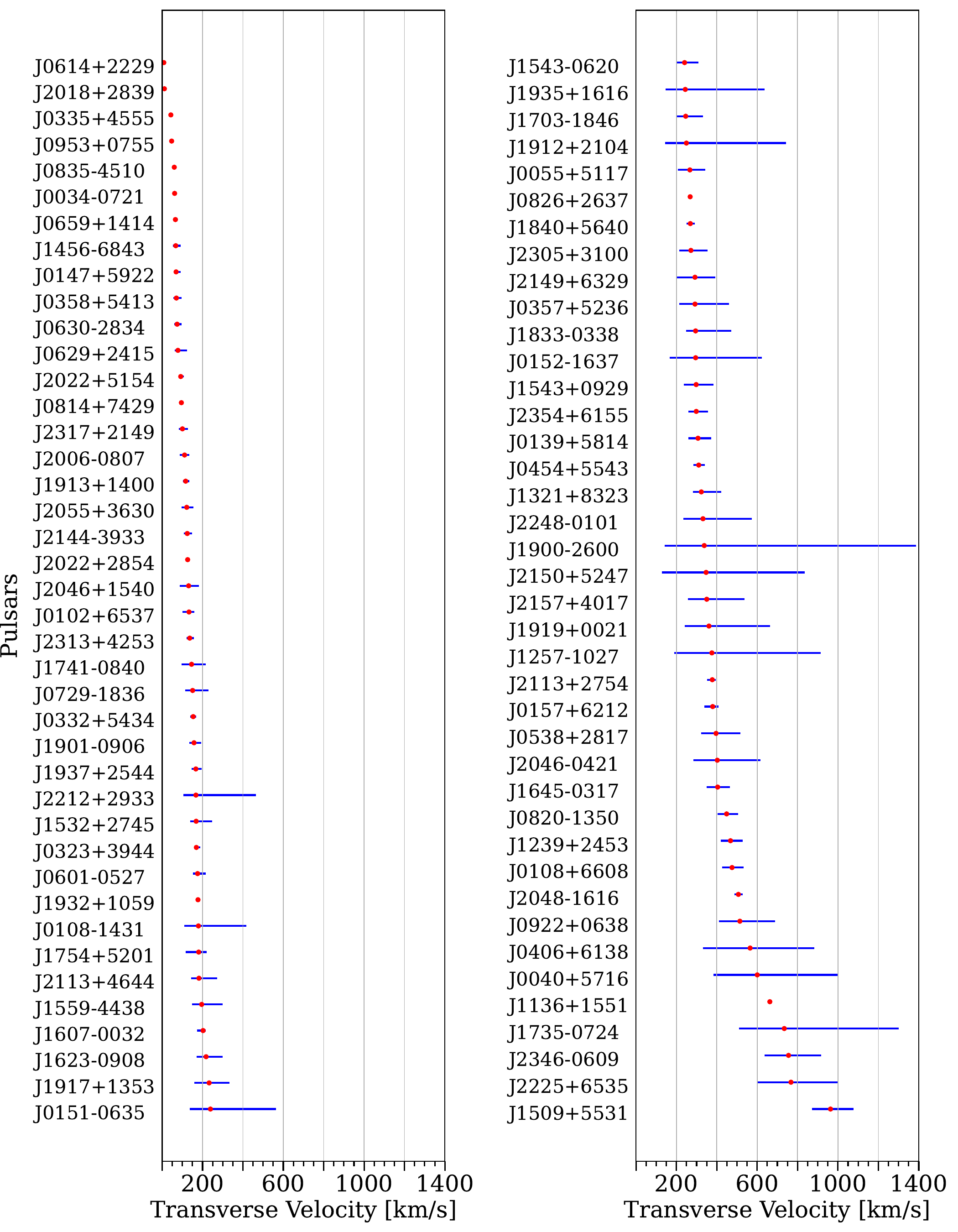}
\caption{Pulsar transverse velocity data from \citet{Willcox:2021kbg}, plotted in ascending order of median velocity. For each pulsar, the median transverse velocity is plotted in red, while the 5\%-95\% confidence interval is shown in blue.}
\label{fig:pulsar_data}
\end{figure}

The velocity observations in this study come from astrometric measurements of isolated pulsars obtained using very long baseline interferometry (VLBI), as these have superior precision and avoid systematic uncertainties associated with other distance measurements \citep{Deller:2019}. The data set is a collection of bootstrapped fits for the parallax, positions, and transverse velocities for 81 pulsars as compiled by \citet{Willcox:2021kbg} (hereafter, W21). 
This data set excludes known binary pulsars, millisecond pulsars, and globular cluster pulsars, so that the observed proper velocities of pulsars in the data set are primarily a consequence of the natal kick received during supernovae. The velocity distribution of these pulsars is shown in Fig.~\ref{fig:pulsar_data}.

\section{Single Stellar Evolution Model}\label{sec:sse}
In order to use pulsar data from W21 to constrain model parameters in MM20, we must obtain predictions for NS kick distributions using the MM20 parameterization. To begin, we obtain these distributions using the Single Stellar Evolution (SSE) simulation mode in the $\texttt{COMPAS}$ rapid population synthesis code \citep{compas:2022ApJS..258...34R, COMPAS:2022, stevenson:2017NatCo...814906S}. This codifies the assumption that the pulsars in the W21 data set originate from single stars. The alternative to this assumption is explored in Section~\ref{sec:bse}.

\subsection{Simulated Kick Distributions}\label{subsec:sse_kick_dist}
We simulate single stars with the supernova kick and remnant mass prescription from MM20, which is implemented as the $\texttt{MULLERMANDEL}$ prescription in $\texttt{COMPAS}$. The NS kick scaling prefactor $v_{\rm ns}$ and the kick scatter parameter $\sigma_{\rm ns}$ are varied across simulations to construct a grid that spans the range \revision{$v_{\rm ns} \in [400, 425, 450, 475, 500, 525, 550, 575, 600, 625, 650, 675, 700]$~km/s and $\sigma_{\rm ns} \in [0.1, 0.2, 0.3, 0.4, 0.5]$, where the ranges were selected based on preliminary likelihood tests.} For each configuration of this {$13 \times 5$} parameter space, we simulate $10^6$ stars starting from zero-age main sequence (ZAMS). The initial mass of each star is drawn from the \citet{Kroupa:2000iv} initial mass function (IMF) with mass limited to the range $[5 M_\odot, 150 M_\odot]$. The lower limit of this range is chosen because lower mass stars do not typically form NSs, even following interactions during binary evolution. The upper limit comes from the typical maximum observed mass of stars. The initial metallicity for all the stars is set to solar metallicity $Z_\odot = 0.0142$~\citep{Asplund:2009fu}, since the pulsar observations to which the NSs will be compared are all in the Milky Way Galaxy. All the other parameters are set to the $\texttt{COMPAS}$ defaults. 

\begin{figure}
\includegraphics[width=\columnwidth]{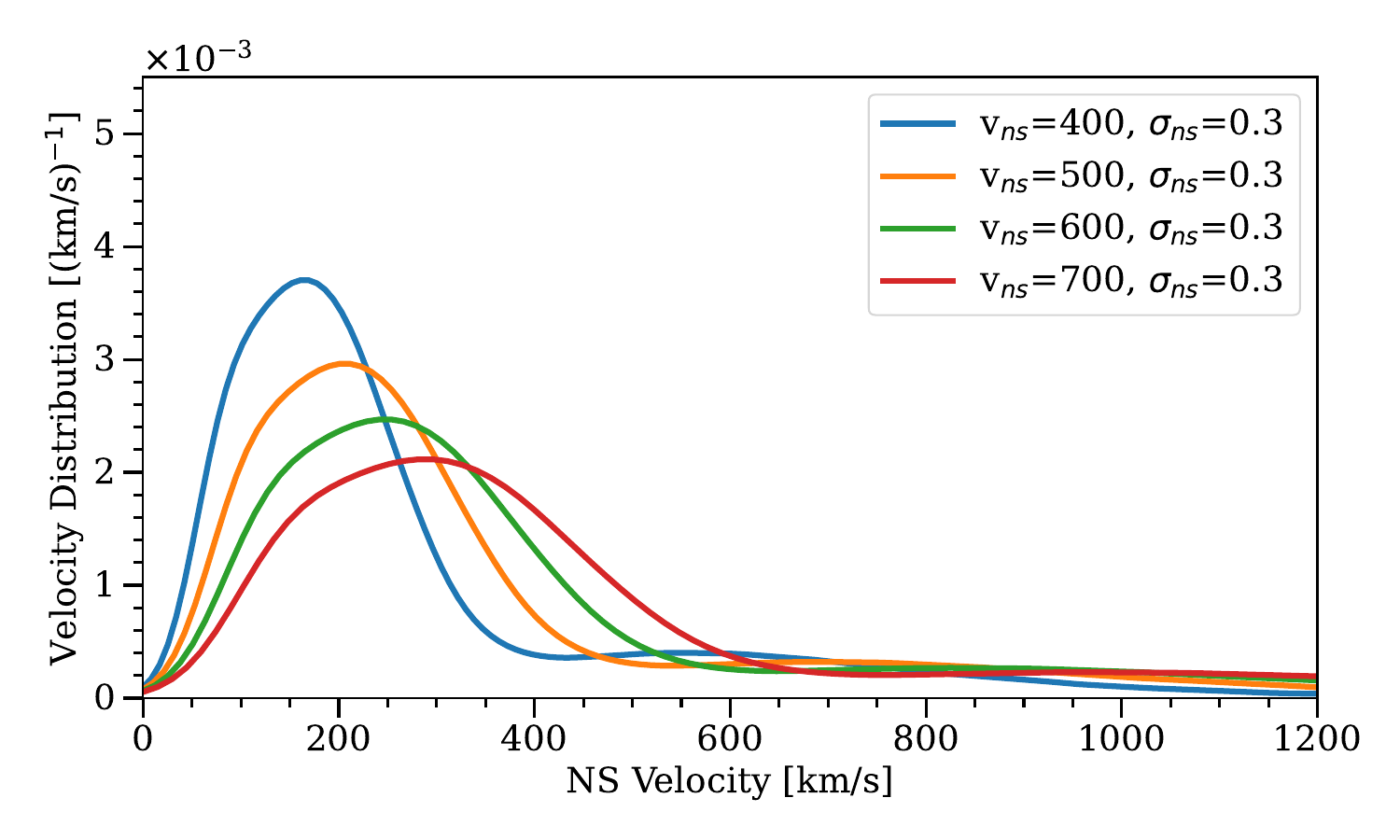}
\includegraphics[width=\columnwidth]{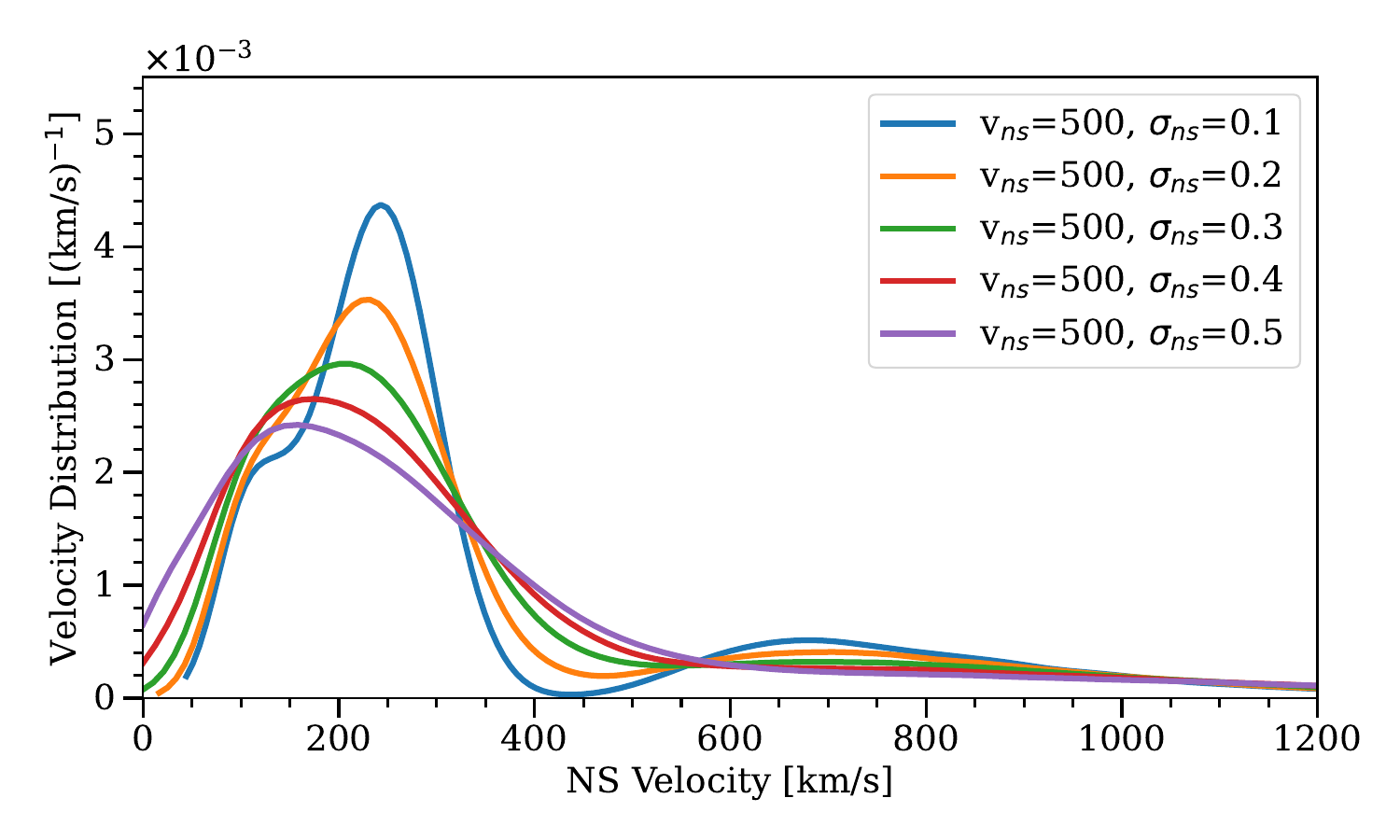}

\caption{NS velocity distributions for various MM20 natal kick prescriptions. The top figure shows the effect of varying $v_{\rm ns}$, while $\sigma_{\rm ns}$ is held constant. The bottom figure shows the effect of varying $\sigma_{\rm ns}$ for a fixed $v_{\rm ns}$. The velocity distributions for all values of ($v_{\rm ns}, \sigma_{\rm ns}$) were obtained \revision{by sampling a population of single stars from the \citet{Kroupa:2000iv} IMF and evolving them using the Single Stellar Evolution (SSE) mode in $\texttt{COMPAS}$.}}
\label{fig:ns_kick_output}
\end{figure}

From the total population of simulated stars, we select the stars that experience CCSNe and leave behind NS remnants to create a set of single pulsars. We ignore the possibility of electron-capture supernovae (ECSNe) under the assumption that single stars experience dredge-up, making it less likely for them to undergo ECSNe~\citep{Miyaji:1980PASJ...32..303M, Podsiadlowski:2003py}; this is consistent with the observational constraints from W21. Note that, since these stars are simulated in isolation, their final velocities are the same as the supernova (SN) natal kicks. Some of the resulting velocity distributions are shown in Fig.~\ref{fig:ns_kick_output}.

When $\sigma_{\rm ns}$ is kept constant, the effect of increasing $v_{\rm ns}$ is to shift the peak of the velocity distribution to the right and broaden it. Thus, on average, NSs receive larger natal kicks for larger values of $v_{\rm ns}$.
The effect of changing $\sigma_{\rm ns}$ while holding $v_{\rm ns}$ constant is slightly more nuanced.  The general behavior is towards a broader spread in kick velocities at higher $\sigma_{\rm ns}$, as expected.  At low values of $\sigma_{\rm ns}$, the kick distributions show an additional feature which is the imprint of the treatment of remnant masses in the MM20 model. The prescription is parameterized such that stars with a CO core mass below $3 M_\odot$ typically lose a smaller fraction of their mass to the supernova ejecta as compared to stars with a CO core mass above $3 M_\odot$ (see Fig. 1 of MM20). As a result, stars with CO cores below $3 M_\odot$ receive smaller kicks than those with CO cores above $3 M_\odot$. This effect can be observed as a dip in the velocity distributions at around $450$ km s$^{-1}$ in the bottom plot of Fig.~\ref{fig:ns_kick_output} for low $\sigma_{\rm ns}$.  This effect is less pronounced for higher values of $\sigma_{\rm ns}$, where the substructure is smoothed out by the larger scatter of kicks. Conversely, large values of $\sigma_{\rm ns}$ increase the spread in kick velocities, but also lower the peak of the distribution.

\subsection{Likelihood Calculation}\label{subsec:likelihood}
Computing the relative likelihoods of configurations with different $v_{\rm ns}$ and $\sigma_{\rm ns}$ requires us to devise a formalism to compare the pulsar distribution from W21 to the simulated kick distributions from Section~\ref{subsec:sse_kick_dist}. We evaluate the likelihood of the observed data given each simulated parameter configuration from the MM20 model. We will call a given natal kick distribution model $\mathcal{M}(\vec{\theta})$, where $\vec{\theta} = \{v_{\rm ns}, \sigma_{\rm ns}\}$ for the MM20 model.

We follow W21 in treating each set of bootstrapped pulsar velocities for pulsar $i$ as a set of samples from the posterior distribution, which, under the assumption that the measurements correspond to uniform priors on pulsar velocities, is proportional to the likelihood of making the pulsar observation $\{d_i\}$ given the particular recorded transverse velocity value $v_{i,k}$. Here, $v_{i,k}$ is the $k$th sample of the $i$th pulsar. Then the likelihood of observing pulsar $\{d_i\}$ given model $\mathcal{M}(\vec{\theta})$ is approximated by a Monte Carlo average over the samples chosen from the posterior,
\begin{equation}
    p(d_i|\mathcal{M}) = \langle p(v_{i,k} | \mathcal{M}) \rangle_k.
\end{equation}
Here, $p(v_{i,k}|\mathcal{M})$ is the probability of drawing a given velocity, which appears in the data set, from model $\mathcal{M}$. 

Finally, the probability of drawing all $N$ pulsars from model $\mathcal{M}$ is
\begin{equation}
    p(d|\mathcal{M}) = \prod_{i=1}^{N} p(d_i|\mathcal{M}).
\end{equation}

\begin{figure}
\includegraphics[width=\linewidth]{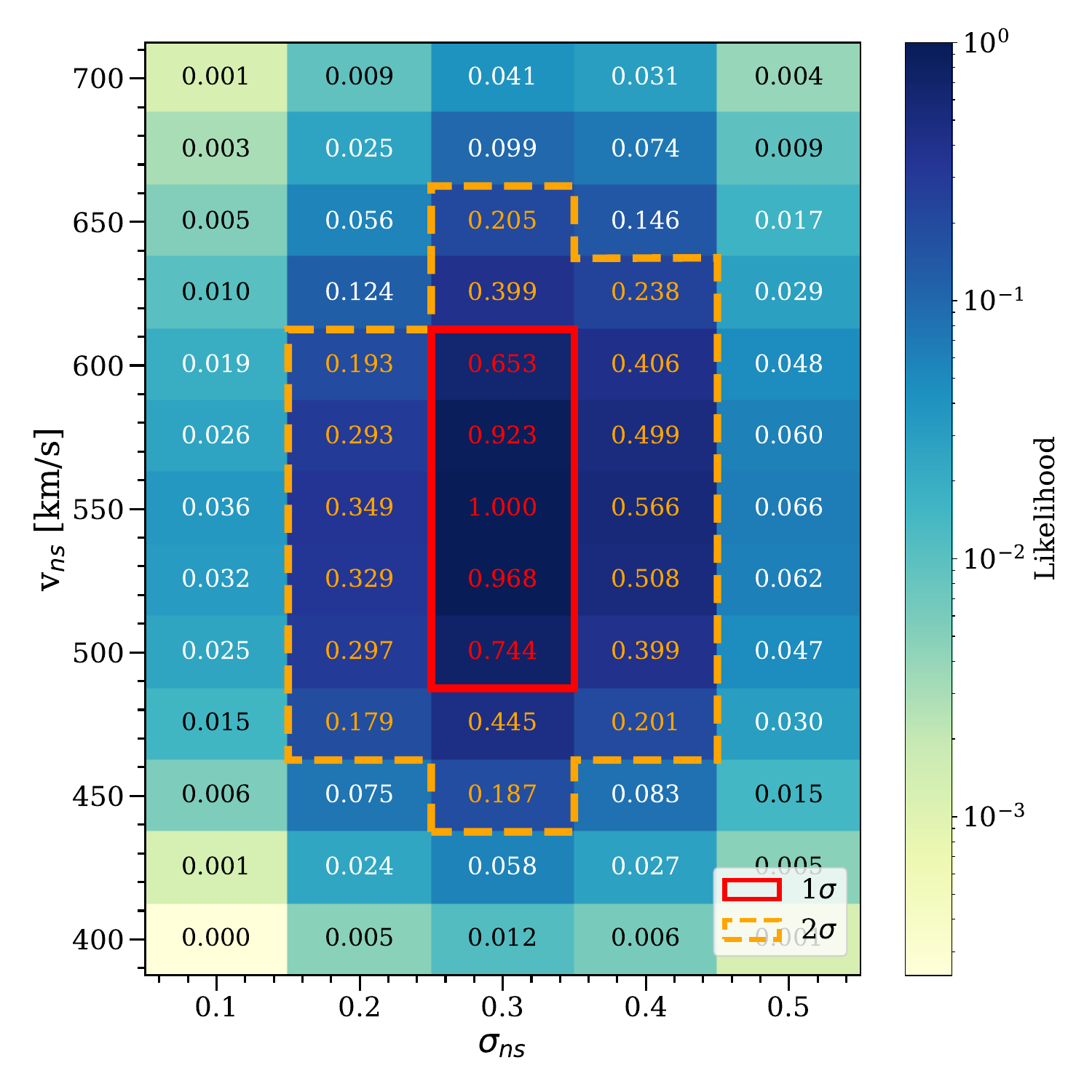}
\caption{Likelihoods for the range of simulated ($v_{\rm ns}, \sigma_{\rm ns}$) models from MM20, calculated for single stellar evolution, and normalized so that the maximum likelihood is unity. The likelihood distribution is fit to a 2D Gaussian, and the parameter space within 1$\sigma$ (2$\sigma$) of the most likely coordinate is highlighted in red (orange).}
\label{fig:model_likelihoods_2d_sse}
\end{figure}

The observed pulsar velocities are transverse 2D velocities, whereas the simulations described in Section~\ref{subsec:sse_kick_dist} produce 3D velocities. In order to compare the simulated velocities to the posteriors, we project the simulated velocities onto the transverse plane assuming isotropic orientation in 3D space, i.e.
\begin{equation}
    v_{\text{2D}} = v_{\text{3D}} \sin \theta,
\label{eq:2d_projection}
\end{equation}
where \revision{$\theta$ is sampled such that $p(\theta) = \sin \theta$.}  The resulting likelihoods for the set of simulated MM20 models are shown in Fig.~\ref{fig:model_likelihoods_2d_sse}.

\subsection{Best-Fit Parameters}\label{subsec:best_fit_sse}
Among the simulated models, the likelihood peaks for the $v_{\rm ns}=550$~km/s, $\sigma_{\rm ns} = 0.3$ model. However, we only simulate certain values of ($v_{\rm ns},\sigma_{\rm ns}$), and the ``true'' parameter values may lie somewhere in between the discrete models included in this analysis. Therefore, we approximate the likelihood distribution over parameter space as a 2D Gaussian to more precisely estimate the maximum likelihood values of ($v_{\rm ns},\sigma_{\rm ns}$). The resulting parameters are $v_{\rm ns} = 550 \pm 112$ km/s, and $\sigma_{\rm ns} = 0.30 \pm 0.16$, where the quoted uncertainties encompass the 95\% credible intervals under the assumption of flat priors on $v_{\rm ns}$ and $\sigma_{\rm ns}$. 

\begin{figure}
\includegraphics[width=\linewidth]{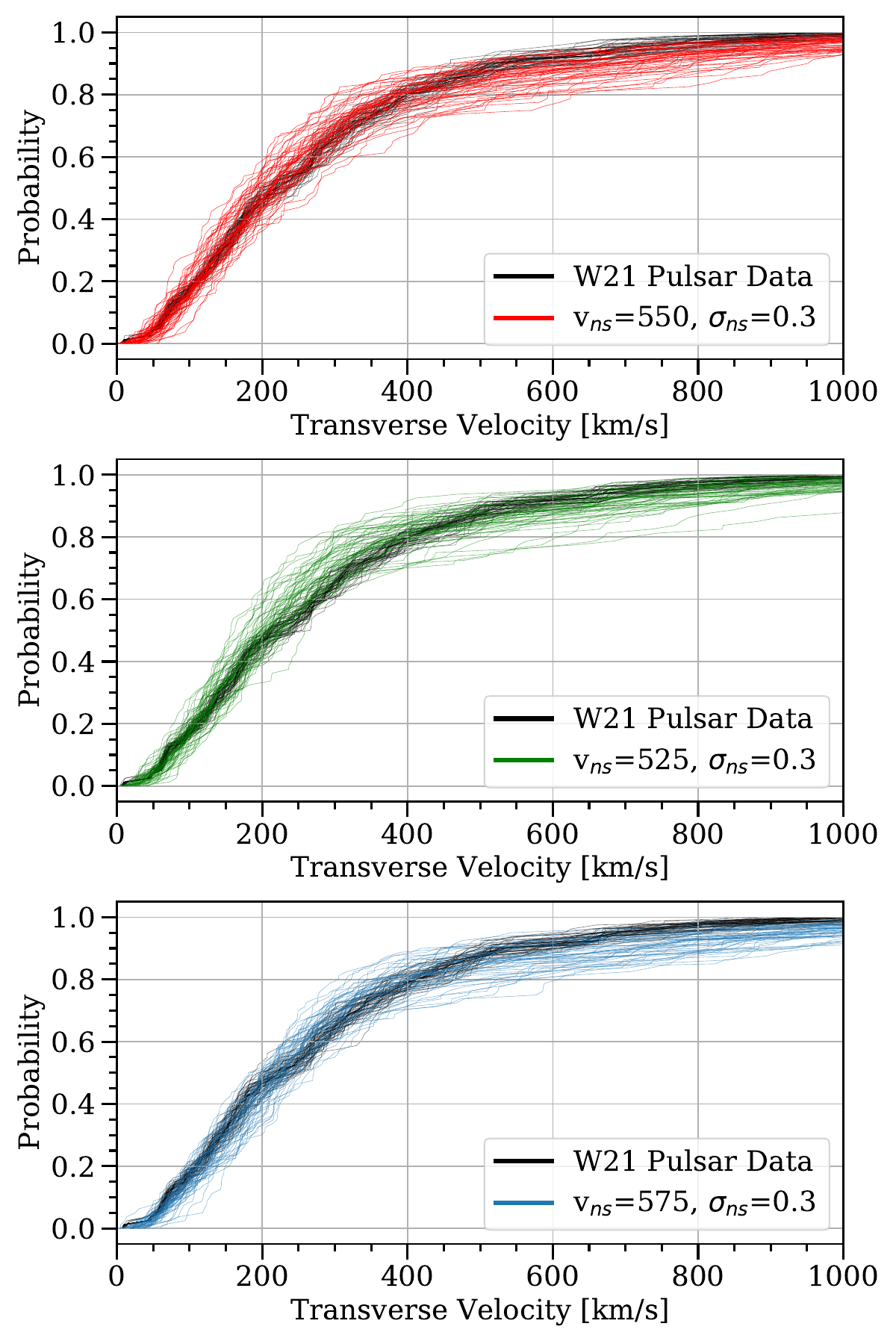}
\caption{CDFs for the transverse velocities of the 3 most likely SSE kick models following the MM20 parameterization (red, green, and blue in order of likelihood), along with pulsar transverse velocity data from \citet{Willcox:2021kbg} (shown in black). The 2D transverse velocities are calculated by taking the final NS velocities from each simulation and projecting them isotropically using Eq. \ref{eq:2d_projection}. All data sets are represented using 50 CDF realizations, with 81 data points each to match the 81 pulsars in the W21 data set.}
\label{fig:model_cdfs}
\end{figure}

It is also important to determine whether the MM20 models in question are consistent with the pulsar data set, since a poor model could be preferred over even worse ones without actually matching the data. We limit this analysis to the top 3 most likely models, hereby identified for brevity as \revision{(550, 0.3), (525, 0.3), and (575, 0.3)}, in order of likelihood. We compare the resulting velocity distributions to the W21 pulsar data visually by studying their cumulative probability distribution functions (CDFs). The complete pulsar data set from \citet{Willcox:2021kbg} is comprised of 81 pulsars, each with a set of equally likely possible inferred transverse velocities. Thus, each observational pulsar data set CDF is constructed from 81 observed velocities, one randomly sampled from each pulsar. Each MM20 model CDF is constructed by sampling 81 NSs from the full simulated single NS catalog and projecting their velocities isotropically onto the sky plane using Eq.~\ref{eq:2d_projection}. The CDFs of the transverse velocities of the best-fit MM20 models, as well as the W21 pulsar data set, are shown in Fig.~\ref{fig:model_cdfs}.
All three MM20 models produce NS transverse velocity CDFs that are visually compatible with the W21 single pulsar population. 

In order to quantitatively check whether these models are consistent with the W21 pulsar data set, we perform a Kolmogorov-Smirnov (KS) test. The formal KS test reveals that all of the three most likely SSE models are consistent with the W21 data set, with p-values ranging from 0.4 to 0.7. Evidently, we fail to reject the null hypothesis that the observed data set could be drawn from the distribution predicted by any of these models. 

\section{Binary Evolution Model}\label{sec:bse}
So far in this work, we have only considered the velocity distributions of initially single NSs. However, the vast majority of massive stars are born in binaries or higher-multiplicity systems \citep{Sana:2012px, MoeDiStefano:2017}. At the same time, several binary evolution codes model SN kick velocities based on single pulsar observations. A priori, we may expect that, for the same assumed natal kick velocity distribution, the binary channel would yield a different distribution of single pulsar velocities.  For example, NSs receiving low natal kicks are more likely to be retained in binaries than those receiving high natal kicks, creating an additional selection effect.  Consequently, there is a risk that inferring the natal kick distribution from observed single pulsars under the assumption of the single-star evolution scenario may lead to misleading results.   As such, it is important to also study the inference on parameters ($v_{\rm ns}$, $\sigma_{\rm ns}$) from binary stellar evolution models.

\subsection{Simulated Kick Distributions}\label{subsec:sse_kick_dist_bse}
The procedure for simulating a kick distribution while accounting for binary interactions is almost identical to Section~\ref{sec:sse}, except we run $\texttt{COMPAS}$ in Binary Stellar Evolution (BSE) mode. As before, the supernova kick and remnant mass prescription is defined according to MM20, with the relevant parameters being varied across the ranges $v_{\rm ns} \in [400, 425, 450, 475, 500, 525, 550, 575, 600, 625, 650, 675, 700]$~km/s and $\sigma_{\rm ns} \in [0.1, 0.2, 0.3, 0.4, 0.5]$ km/s. Each simulation consists of $10^6$ binaries, where the mass of the more massive ZAMS star in each binary, i.e. the primary mass ($m_1$), is drawn from a Kroupa IMF in the range $[5, 150] M_\odot$ as before. \revision{All the binary evolution parameters are set to the $\texttt{COMPAS}$ defaults.} The mass of the secondary star ($m_2$) is then chosen so that the mass ratio $q = m_2/m_1$ follows a uniform distribution on $[0.01,1]$ with $m_2 \geq 0.1 M_\odot$. The initial semi-major axis of the binary is sampled from a uniform-in-log distribution on $[0.01, 1000]$ AU. All the binaries are generated with zero initial eccentricity, and at solar metallicity ($Z = Z_\odot=0.0142$). If a binary is disrupted by the first supernova, the companion star is evolved further to account for the possibility that it too experiences a SN and an additional kick.  

Unlike the SSE scenario, we allow for ECSNe in the binary evolution scenario, as Roche-lobe overflow in a binary system can suppress dredge-up and allow for electron-capture~\citep{Podsiadlowski:2003py, Ibeling:2013bm, DallOsso:2013tnx, Poelarends:2017dua}.  In $\texttt{COMPAS}$, stars explode in ECSNe if they  have helium core masses between 1.6 and 2.25 $M_\odot$ at the base of the asymptotic giant branch, lose their hydrogen envelopes through mass transfer, and reach a carbon-oxygen core mass of 1.38 $M_\odot$ \citep{compas:2022ApJS..258...34R}.  Furthermore, $\texttt{COMPAS}$ assumes that whenever already stripped naked helium stars overflow their Roche lobes after the helium main sequence (case BB mass transfer), they lose their entire helium envelopes but none of their carbon-oxygen core mass and explode in ultra-stripped supernovae (USSNe; \citealt{Tauris:2015}). \revision{However, the amount of mass removed from the donor during case BB mass transfer is uncertain \citep{Tauris:2015, Laplace:2020hum}, and some helium envelope may be retained at the time of the USSN \citep{Yao:2020ley}.}   ECSN and USSN natal kicks follow the same prescription as other NS natal kicks in the MM20 model.

As in the SSE scenario, we select only those stars that have become unbound from their binary companion and ended their evolution as NSs in order to recover the single pulsar population. 
Note that the final velocities of single pulsars that originate in binaries may be different from their natal kicks. This is because pre-supernova orbital velocities contribute to the speeds of ejected pulsars, while the speeds of pulsars formed from secondaries may be impacted second-hand by both asymmetric natal kicks and symmetric mass loss (\citealt{Blaauw:1961} kicks) experienced by the primaries. We do not explicitly select pulsars based on their spin periods in the simulated population, even though millisecond pulsars are excluded from the single-pulsar data set.  However, first-born pulsars in binaries disrupted by the second supernova represent only $\sim$1\% of the total simulated pulsar population, and only a fraction of these are likely to be millisecond pulsars, so this choice is unlikely to impact our conclusions. 

\begin{figure}
\includegraphics[clip,width=\columnwidth]{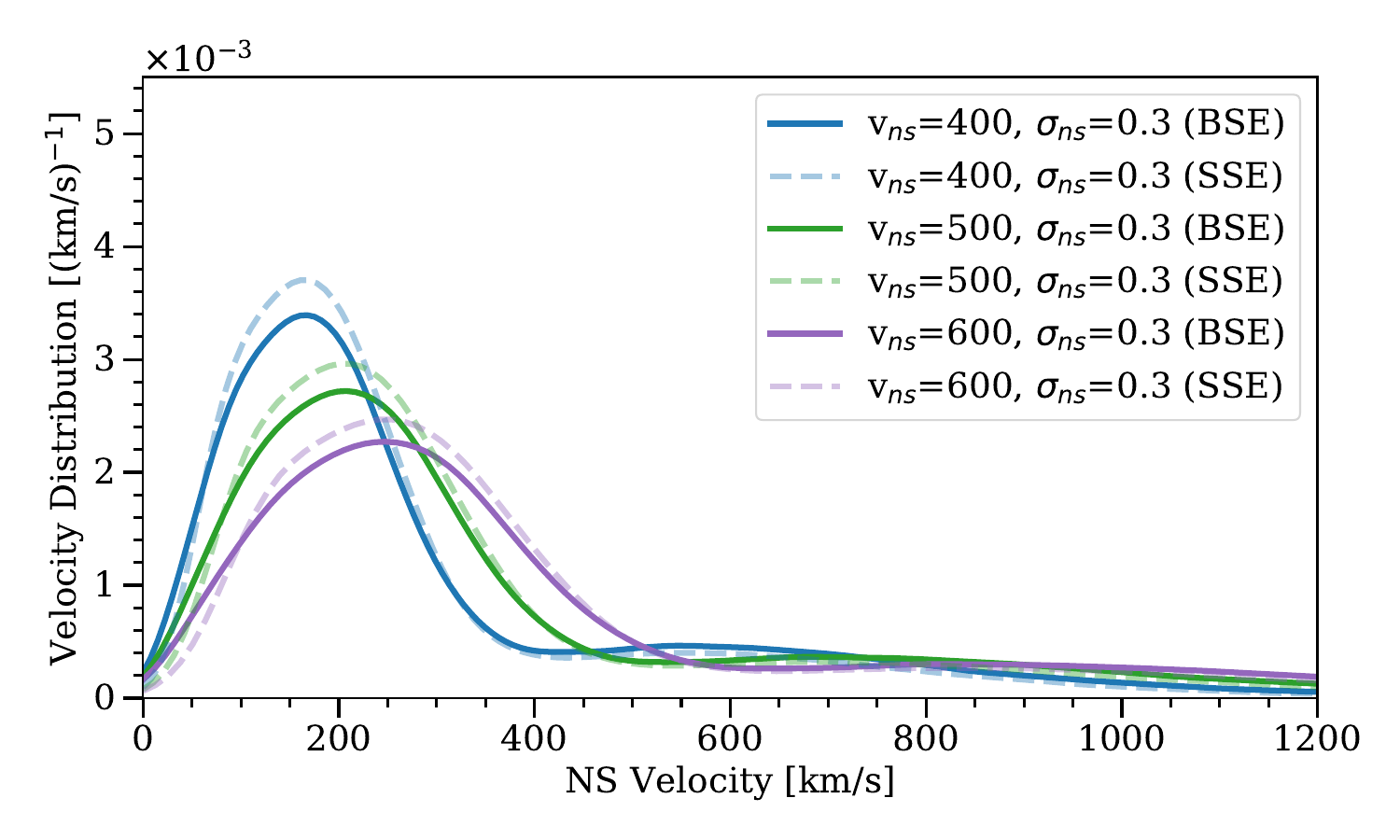}
\includegraphics[clip,width=\columnwidth]{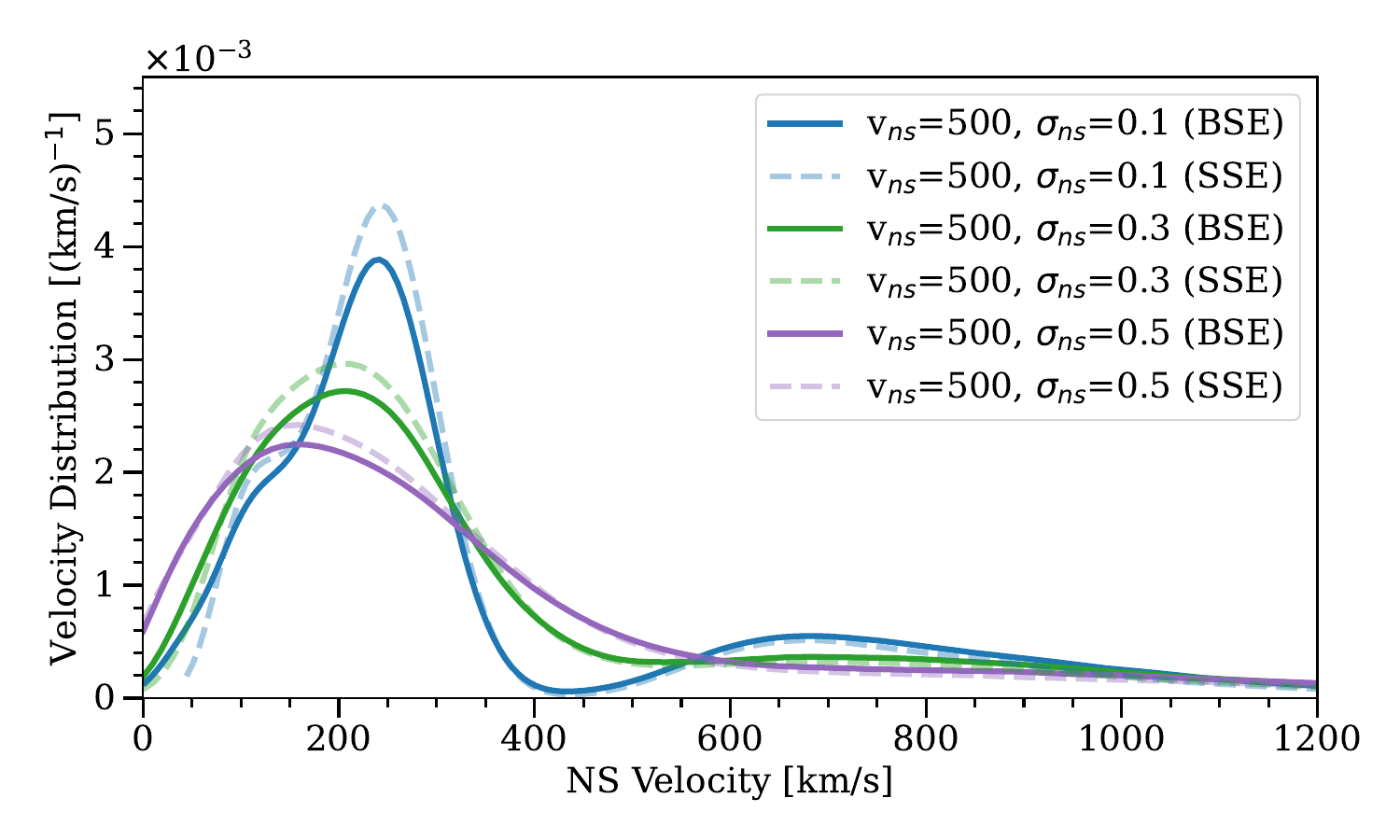}
\caption{NS velocity distributions for some of the configurations in ($v_{\rm ns}, \sigma_{\rm ns}$) space from the MM20 model. The solid (dashed) lines correspond to distributions simulated via Binary (Single) Stellar Evolution.}
\label{fig:ns_kick_output_bse}
\end{figure}

Some examples of the resulting velocity distributions, along with the corresponding curves from the SSE models, are shown in Fig.~\ref{fig:ns_kick_output_bse}.
The modelled velocities of single pulsars produced through binary evolution are mostly similar to those arising from single-star evolution, with a few notable differences. \revision{As noted above, the BSE scenario includes NSs ejected by ECSNe and USSNe, as well as NSs ejected by SN events experienced by their binary companions. These stars typically have lower final velocities than NSs from the single evolution scenario, all of which receive kicks due to CCSNe only. As such, we see a larger number of lower-velocity pulsars in the BSE distributions when compared to SSE.  Furthermore,} we expect the velocities of all ejected stars in the BSE scenario to have a larger statistical spread because of the stochastic addition of orbital velocities to SN kicks.

\subsection{Likelihood Calculation}\label{subsec:likelihood_bse}
\begin{figure}
\includegraphics[width=\linewidth]{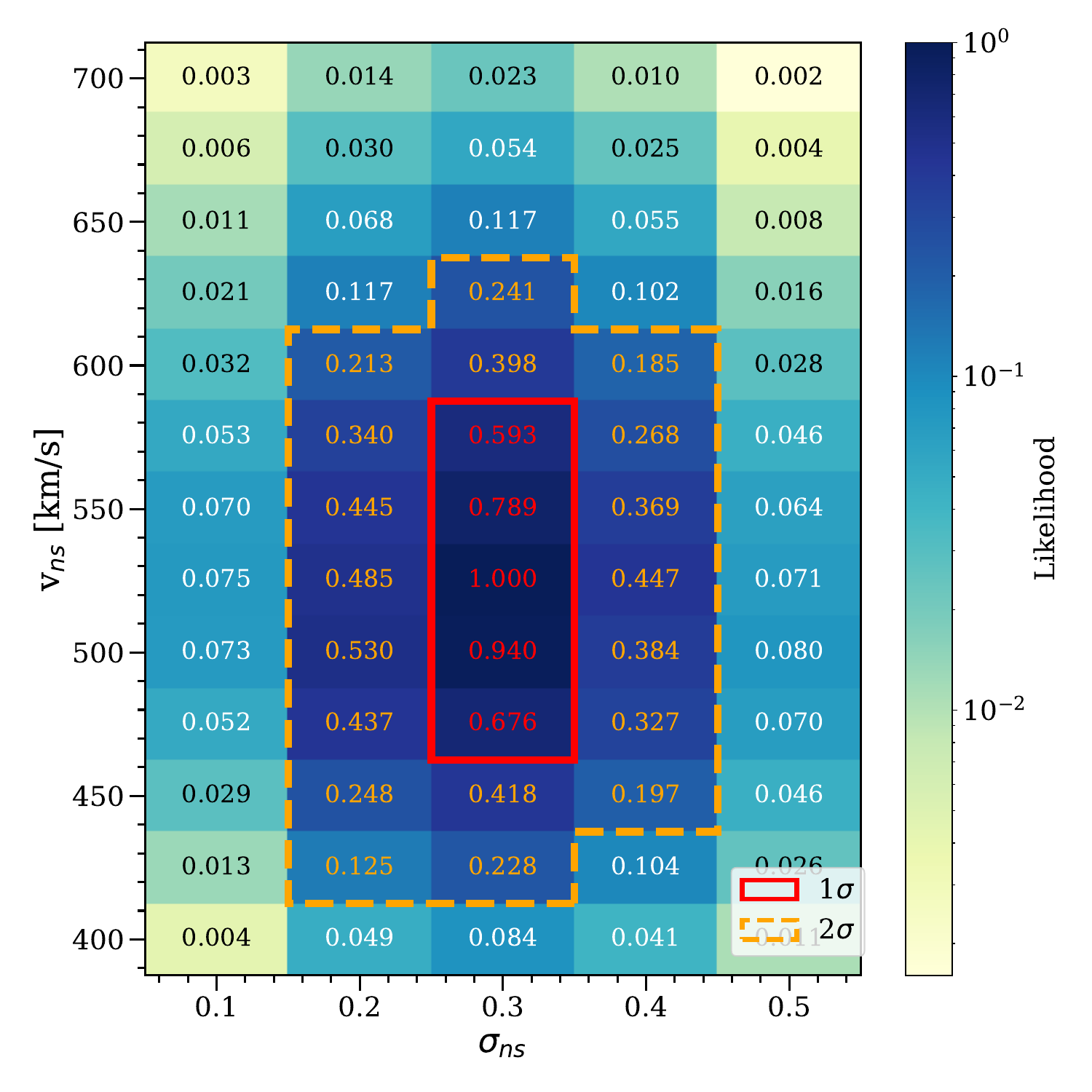}
\caption{Likelihoods for the range of simulated ($v_{\rm ns}, \sigma_{\rm ns}$) models from MM20, calculated for binary evolution, and normalized so that the maximum likelihood is unity. The parameter space within 1$\sigma$ (2$\sigma$) of the most likely model is highlighted in red (orange).}
\label{fig:model_likelihoods_2d_bse}
\end{figure}

We project the simulated pulsar velocity distributions onto the sky plane using Eq. \ref{eq:2d_projection}, and then compute the likelihood for each BSE model parameter configuration as described in Section~\ref{subsec:likelihood}. The resulting likelihoods are shown in Fig.~\ref{fig:model_likelihoods_2d_bse}.

As expected from the similarity of single-pulsar velocities predicted by single and binary evolution channels (see Fig.~\ref{fig:ns_kick_output_bse}), the introduction of binary interactions does not significantly shift the likelihood of the underlying natal kick prescriptions. \revision{Most of the disrupted binaries that would lead to single pulsars are wide, and their orbital velocities are very low compared to the SN natal kick.} As a result, the final velocities of the ejected NSs are primarily set by the natal kicks, with only a small additional spread due to initial orbital velocity. Compared to the SSE scenario, the presence of additional low velocity pulsars and higher scatter due to binary interactions lead to slightly lower preferred values of $v_\text{ns}$ and $\sigma_\text{ns}$. Despite these differences, it is worth noting that the preferred models in the BSE scenario are largely the same as in the SSE scenario, with the \revision{(550, 0.3), (525, 0.3) and (575, 0.3)} models all remaining within 1$\sigma$ of the most likely configuration.
\begin{figure}
\includegraphics[width=\linewidth]{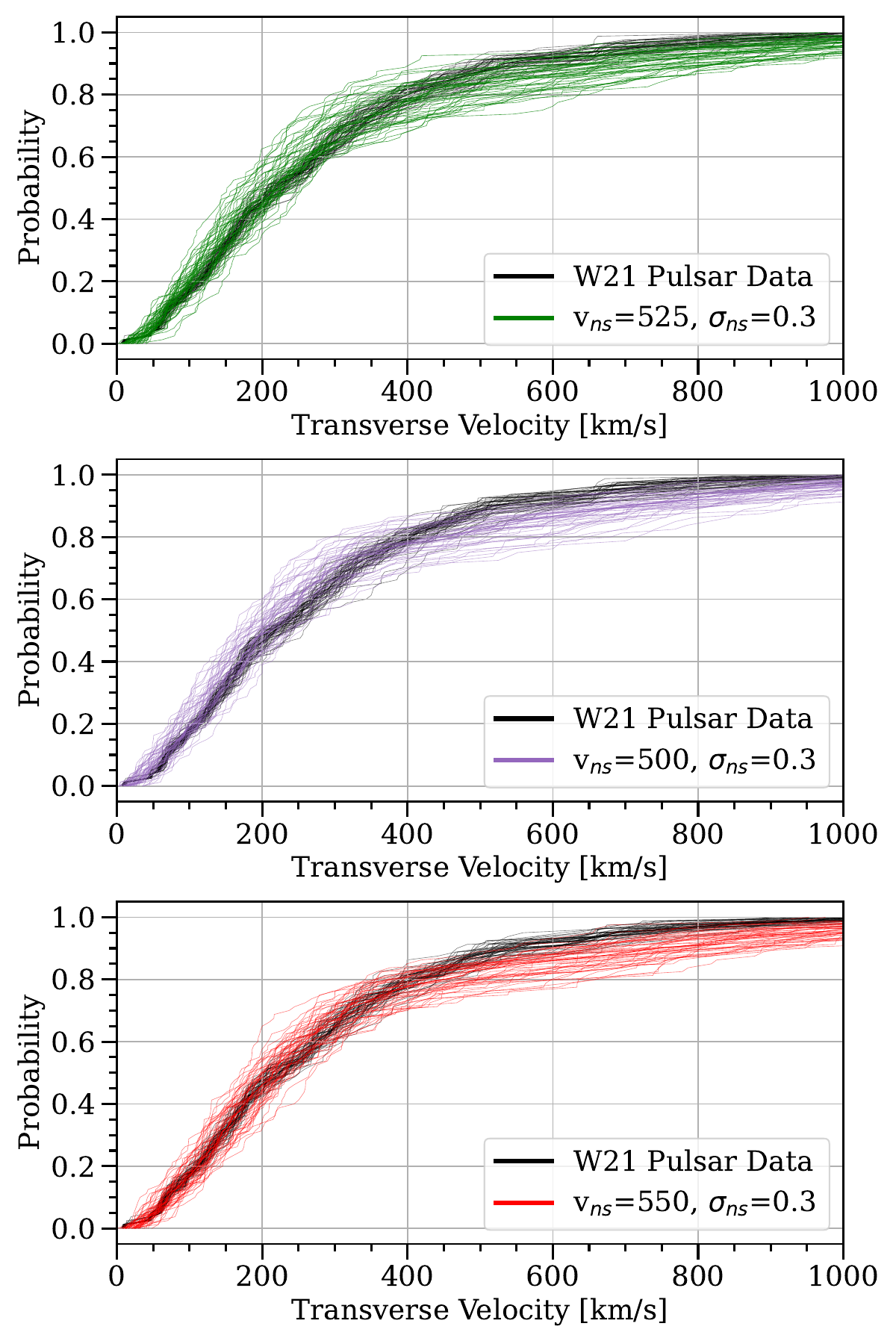}
\caption{CDFs of the single pulsar transverse velocities from the 3 most likely MM20 natal kick models, simulated using BSE, along with pulsar transverse velocity data from \citet{Willcox:2021kbg}. The transverse velocities for the simulated models were calculated using Eq.~\ref{eq:2d_projection}, and all data sets are represented using 50 CDF realizations with 81 data points each.}
\label{fig:model_cdfs_bse}
\end{figure}

\subsection{Best-Fit Parameters}\label{subsec:best_fit_bse}
Fitting a 2D Gaussian to the likelihood distribution, we estimate the best fit parameters to be $v_{\rm ns} = 520 \pm 116$ km/s, and $\sigma_{\rm ns} = 0.3 \pm 0.17$, where the the quoted uncertainties encompass the 95\% credible intervals. We confirm from this calculation that the single and binary best-fit models are well within statistical uncertainty of each other. The consistency between the single and binary models justifies the use of single pulsar observations to infer the natal kick distribution of stars evolving both in isolation and in binaries. Furthermore, this consistency suggests that the inferred kick model is unlikely to be impacted by uncertainties in binary evolution models.

To visually confirm that the three most likely BSE MM20 models: \revision{(525, 0.3), (500, 0.3), and (550, 0.3)}, are consistent with the pulsar population, we plot CDFs of their predicted single NS transverse velocities. These CDFs are shown in Fig.~\ref{fig:model_cdfs_bse}, and are computed using the method described in Section~\ref{subsec:best_fit_sse}. Once again, the models produce single NS populations that appear to mostly overlap with the distribution of single pulsars from observational data. 

We also use a formal KS test to compare the transverse velocity distribution obtained using the three most likely BSE simulations to the W21 data. The resulting p-values range from 0.3 to 0.6, which shows that we fail to reject the null hypothesis for any of the three most likely models. Evidently, the single-pulsar population obtained by applying the MM20 natal kick prescription to binary systems is consistent with the W21 data set.  

\section{Discussion}\label{sec:discussion}
Having identified the best MM20 parameters to reconstruct the observed single young pulsar population, we can now investigate some of the implications of this SN model.

\subsection{BNS Detection Rates}\label{subsec:detection}
A natural application of the SN kick model is to study the resulting detection rate of BNS systems by the LIGO/Virgo/KAGRA (LVK) network during the latest O3 science run, so that we may check if the predicted BNS population is consistent with observations. We choose to compare the detection rate rather than the intrinsic merger rate inferred by \citet{LIGOScientific:2021psn}, since the inferred merger rate is very sensitive to the choice of underlying mass distributions, which are generally not consistent with our astrophysical predictions. 
We use the $\texttt{COMPAS}$ population synthesis code to simulate two different DCO populations, each with $2\times 10^7$ initial binaries, such that we can compare their BNS detection rate predictions. 
\begin{enumerate}
\item MM20: 
In the first population, we draw natal kicks from the MM20 prescription with $v_{\rm ns}=520$ km/s and $\sigma_{\rm ns}=0.3$ following the best-fit parameters identified in Section~\ref{subsec:best_fit_bse}. The prescription is consistent across the various SN types: CCSNe, ECSNe, and USSNe. This means that any differences in the kick velocities between SN types emerge only as a consequence of the CO core masses of the progenitors and the NS remnant masses.

\item Maxwellian: 
In the second population, we draw natal kicks from a Maxwell-Boltzmann distribution with the 1D root-mean-square velocity $\sigma_{\text{kick, rms}}$ being determined by the type of SN. For CCSN, we set the 1D root-mean-square velocity to $\sigma_{\text{kick, rms}}= \sigma_{\rm CCSN, rms}=265\, \text{km/s}$, following \citet{Hobbs200510.1111/j.1365-2966.2005.09087.x}. The ECSN and USSN kicks are set to $\sigma_{\text{ECSN, rms}}= \sigma_{\rm USSN, rms}=30\, \text{km/s}$ \citep{Vigna-Gomez:2018dza}. These parameters are chosen because they represent some of the most commonly used SN kick prescriptions. 

\end{enumerate}

The rest of the binary parameters are kept consistent between the two sets of populations, and are identical to the simulations described in Section~\ref{sec:bse}.  Some of the physical assumptions, such as the uncertain physics of mass transfer and common-envelope evolution, may very significantly change our results \citep[e.g.,][]{Broekgaarden:2022}.  However, our primary goal here is to consider specifically the impact of the supernova kick prescription rather than to faithfully reproduce the merging BNS population.

\begin{figure}
\includegraphics[width=\linewidth]{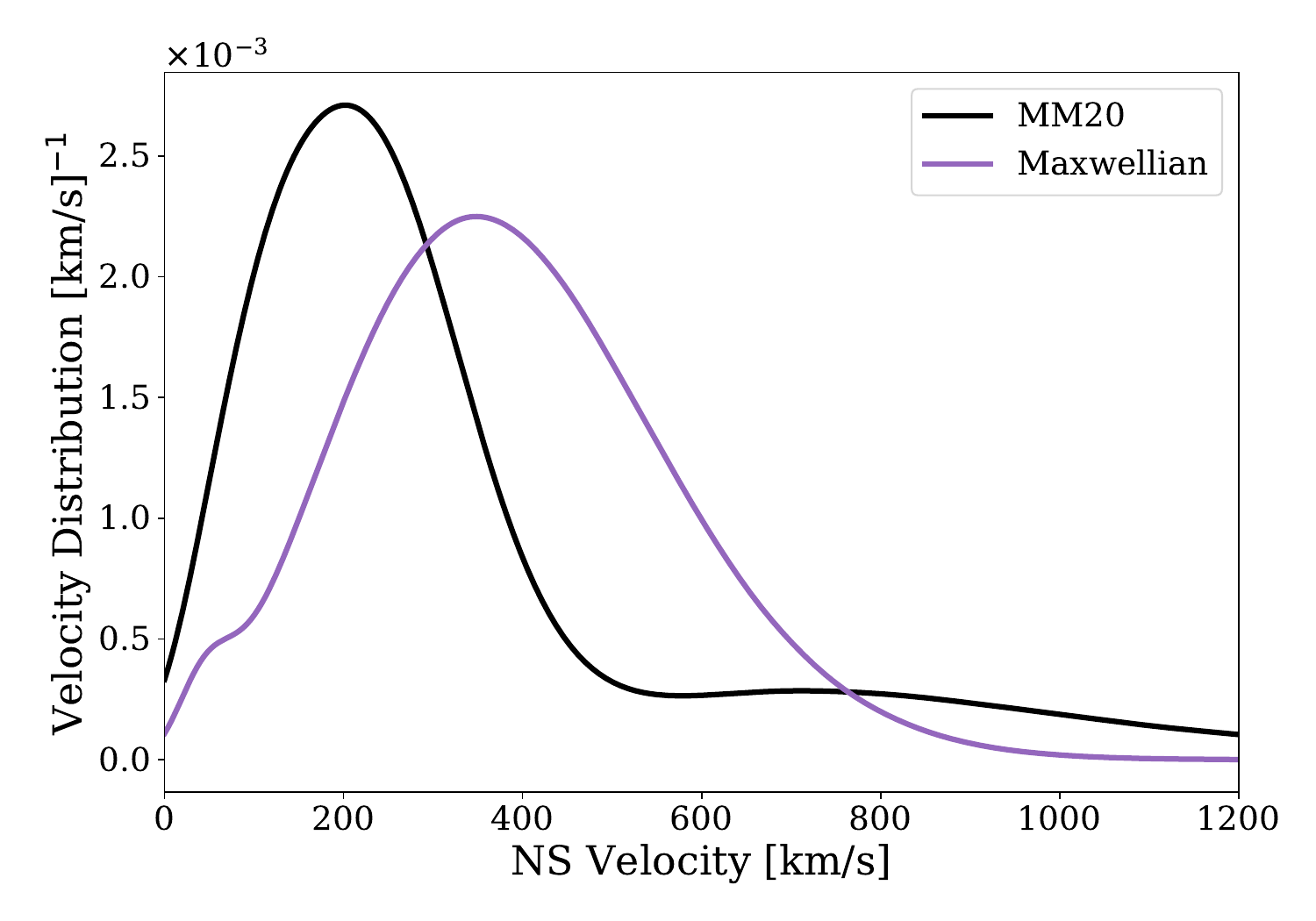}
\caption{Velocity distributions of ejected single NSs, simulated using the following two SN kick prescriptions: the (520, 0.3) MM20 model, and a Maxwellian distribution with $\sigma_{\rm CCSN, rms}=265 \text{km/s}$~\citep{Hobbs200510.1111/j.1365-2966.2005.09087.x} and $\sigma_{\text{ECSN, rms}}= \sigma_{\rm USSN, rms}=30 \text{km/s}$~\citep{Vigna-Gomez:2018dza}.}
\label{fig:hobbs_comparison}
\end{figure}

One change from previous sections is in the metallicity distribution, where instead of evolving all the binaries at solar metallicity, we sample from a log-uniform distribution of metallicities in the range $[0.0001, 0.03]$. This is done to capture the evolution of the BNS formation and merger rate density evolution over cosmic history. The distributions of single NS velocities in the two simulated populations are shown in Fig.~\ref{fig:hobbs_comparison}.

The method for calculating the BNS detection rate of each simulated population follows the procedure described in \citet{Neijssel:2019irh} and \citet{Broekgaarden:2021iew}. 
Our simulation accounts for only a fraction of the total stellar mass in the universe, since we ignored primary stars with mass below $5 M_\odot$ and single stars. Therefore, we first calculate the total star forming mass represented by our simulation at each value of metallicity $Z_i$. The total number of BNS systems formed in the simulation can then be normalized to obtain a BNS formation rate per unit star forming mass (SFM), i.e. $\text{d}N_{\text{form}}/\text{d}M_{\text{SFM}}$. We can then obtain the formation rate of BNS systems at a given metallicity $Z_i$ per unit star forming mass, with component masses $m_1, m_2$ and delay time $t_\text{delay}$, defined as
\begin{equation}
\begin{multlined}
R_\text{BNS}(Z_i, t_\text{delay}, m_1, m_2) = \\
    \frac{\text{d}^4 N_\text{form}}{\text{d}M_\text{SFM} \text{d} t_\text{delay} \text{d}m_1, \text{d}m_2}(Z_i, t_\text{delay}, m_1, m_2).
\end{multlined}
\end{equation}

Here, $t_\text{delay}$ refers to the time between formation of the ZAMS binary and merger of the BNS system.
We can then calculate the BNS merger rate by integrating $R_\text{BNS}$ over metallicity and time, such that the merger rate density (per unit time per unit comoving volume per unit component masses) at a time $t_\text{m}$, for binaries with masses $m_1$ and $m_2$, is given by
\begin{equation}
\begin{multlined}
R_\text{m}(t_\text{m}, m_1, m_2) =  \\
\int \text{d}Z_i \int_0^{t_\text{m}} \text{d}t_\text{delay} \quad \text{MSSFR}(Z_i, t_\text{form}) \times \\
 R_\text{BNS}(Z_i, t_\text{delay}, m_1, m_2).
\end{multlined}
\end{equation}
Here, MSSFR$(Z_i, t_\text{form})$ = MSSFR$(Z_i, z_\text{form})$ is the metallicity-specific star formation rate (SFR) and the formation time is $t_\text{form} = t_\text{m} - t_\text{delay}$. The MSSFR is given by
\begin{equation}
\begin{multlined}
    \text{MSSFR}(Z_i, z_\text{form}) = \frac{\text{d}^3 M_\text{SFR}}{\text{d}t_\text{s} \text{d}V_\text{c} \text{d} Z_i}(z_\text{form}) \\
    = \frac{\text{d}^2 M_\text{SFR}}{\text{d}t_\text{s} \text{d}V_\text{c}}(z_\text{form}) \times \frac{\text{d}P}{\text{d}Z_i}(z_\text{form}),
\end{multlined}\label{eq:mssfr}
\end{equation}
where $t_\text{s}$ is the time in the source frame of the merger, and $V_\text{c}$ is the comoving volume. 

The first term in Eq. \ref{eq:mssfr} corresponds to the cosmological star forming rate. We use the preferred SFR model from \citet{Neijssel:2019irh}, which is based on the phenomenological form developed by \citet{Madau:2014bja}. The second term is a metallicity density function, which is also set to the fiducial model from \citet{Neijssel:2019irh}, i.e. a log-normal distribution in metallicity whose form follows  \citet{Langer:2005hu}. Note that, although the MSSFR choice significantly impacts the merger rate density, this impact is generally least important for BNS merger rates \citep{Chruslinska:2019,Neijssel:2019irh,Broekgaarden:2022}.  We follow these prescriptions across all calculations presented in this work.

As a final step, we convolve the merger rate with observational selection effects as described in \citet{Barrett:2017fcw} to estimate the gravitational-wave detection rates.  We use the LIGO O3 sensitivity curve and a signal to noise (SNR) threshold of 8 in a single detector as proxy for detectability by the network.

The resulting detection rates for the two models are shown in Table \ref{table:detection_rates}. The errors on the quoted value represent the 95\% confidence interval from 500 bootstrapped calculations.  As mentioned earlier, these statistical errors are much smaller than the possible systematic errors stemming from uncertain assumptions about other physics that we kept fixed across the models.   So far, there have been at least two confirmed BNS detections identified by the LVK collaboration over an observing span of less than two years \citep{LIGOScientific:2021psn}. The detection rate computed using the MM20 (and Maxwellian) natal kick prescription and presented in Table \ref{table:detection_rates} falls short of the current LVK detection rate.  This points to the need for other improvements to the binary evolution model, such as in the treatment of the common-envelope phase \citep[e.g.,][]{HiraiMandel:2022}; however, our goal here is to focus on the impact of natal kick prescriptions.

\begin{table}
\centering
\begin{tabular}{ c  c }
\hline \hline  
Natal Kick Model   & BNS Detection Rate   \\ 
                   & (yr$^{-1}$)  \\ 
\hline \hline
MM20  &  0.09 $\pm$ 0.01  \\ 
\hline
Maxwellian        &   0.14 $\pm$ 0.02\\ 
\hline
\hline
\end{tabular}
\caption{Predicted O3 detection rates of BNSs in populations evolved using two different SN kick prescriptions, as described at the beginning of Section~\ref{subsec:detection}. The uncertainties represent the 95\%
confidence intervals. The MSSFR prescription used in the calculation is from \citet{Neijssel:2019irh}.}
      \label{table:detection_rates}
\end{table}

For the same set of assumptions about SFR and cosmic metallicity evolution, the local BNS detection rate given by the MM20 model is slightly lower than the Maxwellian model. When studying the underlying BNS populations, we find that of all SN events where the remnant is a NS, the MM20 model disrupts a larger fraction of binaries than the Maxwellian model.  This is perhaps counter-intuitive, as Fig.~\ref{fig:hobbs_comparison} shows that the ejected NS velocities predicted by the MM20 model are systematically lower than those obtained from the Maxwellian distribution.  The difference in the fraction of disrupted binaries can be understood by separately considering different types of SN events. 

While the MM20 model favors lower CCSN kicks than the Maxwellian model, these kicks are generally high enough in both cases to eject most NSs from their binary systems. Indeed, CCSNe eject $\sim 98.1\%$ of the NSs they are applied to in the Maxwellian population, and $\sim 97.7\%$ NSs in the MM20 population. 
The ECSN kick distributions are largely consistent between the two populations, and we find that ECSNe eject NSs roughly $52.4\%$ and $56.4\%$ of the time in the Maxwellian and MM20 populations, respectively. The main difference between the two populations arises in the case of USSNe. 
We find that the Maxwellian USSNe kicks eject $\sim 1.4\%$ of NSs they are applied to, while the MM20 USSNe eject $\sim 39\%$. In summary, the MM20 model has a lower NS ejection rate for CCSNe, but a higher NS ejection rate for ECSNe and USSNe when compared to the Mawellian prescription. \revision{In our simulations, we find that} USSNe are critical in forming merging BNS, with $96\%$ of BNS that would merge within 14 Gyr experiencing a USSN as the second SN in the binary. \revision{This is consistent with~\citet{Tauris:2015}, who conclude that for a BNS system to merge promptly, the second SN must happen in a very close binary where the secondary would have been ultra-stripped by case BB mass transfer.} The greater fraction of disruptions due to larger USSN kicks in the MM20 model  directly translates into a decrease in predicted BNS detection rates.

\subsection{Globular Cluster Retention}\label{subsec:gc_retention}
The NS kick distribution predicted by the MM20 (520, 0.3) model can also be used to estimate the fraction of NSs retained in globular clusters (GCs), which host far more NSs than expected if natal kicks are large given the typical escape velocities of $\sim 50$ km s$^{-1}$ \citep[e.g.,][]{Sigurdsson:2003}. We find that for an escape velocity of 50 km s$^{-1}$, 6.2\% of all the NSs formed in binaries in a GC would remain in the cluster for our preferred model. We can compare this to the retention fraction predicted using the Maxwellian kick model, which is $6.7\%$. These retention fractions follow the trend we observed in Section~\ref{subsec:detection}, where the MM20 kick model results in more NS ejections than the Maxwellian model. 
Among NSs that evolve alone, the MM20 model results in the retention of $\sim 1\%$ of NSs in a cluster, while the Maxwellian model leaves only $\sim 0.2\%$ of single NSs in GCs.

\subsection{Implications}

The MM20 model may overestimate the explosion energies for supernovae with little support by turbulent convection, as is likely the case for ECSNe and USSNe.  Furthermore, those explosions tend to be more clumpy and less unipolar, which
implies that the momentum anisotropy is lower than for classical CCSNe.  Reduced ECSN and USSN kick velocities could be consistent with the reduced explosion energies predicted for USSNe and possibly matching observations of USSN candidates \citep{Suwa:2015} as well as models that predict $\lesssim$ few km s$^{-1}$ ECSN natal kicks \citep{Gessner:2018}.  In fact, given the paucity of direct observational constraints on the natal kicks associated with ECSNe and USSNe, binary survival and globular cluster retention may be strong indicators of the need to reduce these kicks in the MM20 model.  A possible alternative to the MM20 model would draw ECSN kicks from Maxwell-Boltzmann distributions with 1D root-mean-square speeds of 5 km s$^{-1}$ and halve USSN kicks relative to those of CCSNe with the same progenitor core masses and remnant masses.  This would not appreciably impact the velocity distribution of observed single pulsars, but would increase BNS merger rates by up to a factor of 3 by suppressing ECSN and USSN binary disruptions.

On the other hand, even a factor of 3 increase in the predicted detection rate would not bring predictions in line with gravitational-wave observations.  This discrepancy is not unexpected, because the physics of mass transfer, particularly ultra-stripping, is almost certainly incomplete in rapid binary population synthesis models.  For example, these models fail to accurately predict the observed period-eccentricity distribution of Galactic BNS observed as radio pulsars \citep{AndrewsMandel:2019} or their mass distribution \citep{Vigna-Gomez:2018dza}.  Meanwhile, \citet{Schneider:2020vvh} proposed that the structural changes brought on by mass transfer may impact the explodability of a star, and hence its remnant mass and kick.  

\section{Conclusions}\label{sec:conclusions}
The best-fit MM20 SN natal kick parameterization explored in this work has several desirable characteristics. Compared to the most commonly used empirical fits such as the Maxwellian model, the MM20 model produces velocity distributions that are physically motivated by supernova simulations and account for the impact of progenitor and remnant masses on natal kicks while retaining a degree of stochasticity. Our kick model produces single NS populations that are consistent with single pulsar observations while fitting only two free parameters.

We used the transverse velocities of single NSs as the only observational constraint for inferring natal kick parameters.  The velocities of binaries that retain NSs, such as black widow and redback pulsars and NS low-mass X-ray binaries, could provide additional constraints, but the inference will be more sensitive to other assumptions about binary physics, such as mass transfer. Additionally, while we limited this work to kicks received by NSs, it would also be interesting to constrain black-hole natal kicks and study how they affect mergers involving one or more black holes.  We leave this work for future studies.

\section*{Acknowledgements}

V.K. thanks Reinhold Wilcox for providing access to the pulsar data set, and for his continuous assistance with the simulations. The authors thank Bore Gao and Philipp Podsiadlowski for helpful discussions. V.K. and E.B. are supported by NSF Grants No. AST-2006538, PHY-2207502, PHY-090003 and PHY20043, and NASA Grants No. 19-ATP19-0051, 20-LPS20- 0011 and 21-ATP21-0010. 
I.M.~acknowledges support from the Australian Research Council Centre of Excellence for Gravitational  Wave  Discovery  (OzGrav), through project number CE17010004. I.M.~is a recipient of the Australian Research Council Future Fellowship FT190100574.  Part of this work was performed at the Aspen Center for Physics, which is supported by National Science Foundation grant PHY-1607611.  I.M.'s participation at the Aspen Center for Physics was partially supported by the Simons Foundation.  This research was supported in part by the National Science Foundation under Grant No.~NSF PHY-1748958.
The authors also acknowledge the Texas Advanced Computing Center (TACC) at The University of Texas at Austin for providing HPC resources that have contributed to the research results reported within this paper. URL: \url{http://www.tacc.utexas.edu}~\citep{10.1145/3311790.3396656}.

\section*{Data Availability}
The data underlying this article will be shared on reasonable request to the main author.

\bibliographystyle{mnras}
\bibliography{refs} 

\bsp
\label{lastpage}
\end{document}